\UseRawInputEncoding
\documentclass[aps,floatfix,amsmath,nofootinbib,amssymb,superscriptaddress]{revtex4}

\usepackage{overpic}
\usepackage{amssymb}
\usepackage{indentfirst}
\usepackage{feynmf}   
\usepackage{slashed}  
\usepackage{cases}
\usepackage{color}
\usepackage{multirow}
\usepackage{epstopdf}
\usepackage{graphicx,color,bm}
\usepackage{epstopdf}

\usepackage[colorlinks,
            citecolor=blue,
            anchorcolor=red,
            menucolor=red,
            linkcolor=red,
            filecolor=red,
            runcolor=red,
            urlcolor=blue,
            frenchlinks=red]{hyperref}

\begin{document}

\title{$\sin (\phi_\Lambda-\phi_S)$ azimuthal asymmetry in transversely polarized $\Lambda$ production in SIDIS within TMD factorization at EIC}
\author{Hui Li}
\affiliation{School of Physics, Southeast University, Nanjing 211189, China}
\author{Xiaoyu Wang}
\email{xiaoyuwang@zzu.edu.cn}
\affiliation{School of Physics and Microelectronics, Zhengzhou University, Zhengzhou, Henan 450001, China}
\author{Zhun Lu}
\email{zhunlu@seu.edu.cn}
\affiliation{School of Physics, Southeast University, Nanjing 211189, China}

\begin{abstract}

We investigate the $\sin (\phi_\Lambda-\phi_S)$ single-spin asymmetry in the transversely polarized $\Lambda$ production in semi-inclusive deeply inelastic scattering process within the framework of the transverse momentum dependent~(TMD) factorization.
The asymmetry is contributed by the convolution of the polarizing TMD fragmentation function  $D_{1T}^\perp$ of the $\Lambda$ hyperon and the unpolarized TMD distribution function $f_1$ of the target nucleon.
We apply two sets of $D_{1T}^{\perp,\Lambda^\uparrow/q}$, the spectator diquark model result and available parametrization, to numerically estimate the $\sin (\phi_\Lambda-\phi_S)$ asymmetry at the kinematical region of electron ion collider~(EIC).
We take into account the TMD evolution effect in order to include the scale dependence of the TMD distribution functions and fragmentation functions.
It is found that different sets of $D_{1T}^{\perp,\Lambda^\uparrow/q}$ lead to different ${\sin (\phi_\Lambda-\phi_S)}$ asymmetry, particularly in sign.
Future measurements on the $\sin (\phi_\Lambda-\phi_S)$ asymmetry with high precision at EIC can provide important cross check on the available $\Lambda$ polarizing fragmentation functions as well as constrain them more stringently.
\end{abstract}

\maketitle

\section{Introduction}

Understanding the internal partonic structure of hadrons and the fragmentation mechanism of partons are among the main goals in QCD and high energy physics.
Once the production of a polarized lambda hyperon from unpolarized $pp$ collisions has been observed~\cite{Lesnik:1975my,Bunce:1976yb}, it has become a long-standing challenge~\cite{Kane:1978nd,Dharmaratna:1996xd} in QCD spin physics since such polarization should be small in leading twist in the collinear picture~\cite{Kane:1978nd}.
The traditional theory expects the single spin asymmetries should be forbidden in the partonic level and the averaged polarization of $\Lambda$ should be zero~\cite{Dharmaratna:1996xd}.
Thus, the production of a transversely polarized $\Lambda$ provides an opportunity not only to study the spin structure~\cite{Ma:2001rm} but also the fragmentation mechanism~\cite{Jaffe:1996wp,Burkardt:1993zh,Kanazawa:2015jxa,Boer:2010ya,Anselmino:2001js} of partons.

After introducing the intrinsic transverse momentum into the collinear picture, the transverse single spin asymmetry can originate from the correlation of the transverse motion of the parton and the transverse spin of the hadron.
It is suggested~\cite{Anselmino:2000vs} that a polarizing fragmentation function~(FF)~\cite{Mulders:1995dh}, denoted by $D_{1T}^\perp(x,\bm k_T^2)$, can account for the polarization of the $\Lambda$ production.
$D_{1T}^\perp$ is a time-reversal-odd~(T-odd) and transverse momentum dependent~(TMD) FF, which describes the fragmentation of an unpolarized quark to a transversely polarized hadron, and reflects the correlation of the transverse spin of the produced $\Lambda$ and the transverse momentum of the parent quark.
Sometimes it is viewed as the analog of the Sivers function~\cite{Sivers:1989cc,Sivers:1990fh}, which is a T-odd TMD parton distribution function (PDF) describing the asymmetric density of unpolarized quarks inside a transversely polarized nucleon.
Furthermore, $D_{1T}^\perp$ may play an important role in the spontaneous polarization, such as the process $q\rightarrow \Lambda^{\uparrow}X$~\cite{Boer:2009uc}.
Thus, the study on the production of polarized $\Lambda$ could also provide the information on the spin structure of the hyperon.
This is intriguing since the $\Lambda$ hyperon can not serve as a target in high energy scattering processes.

Experimentally, the single inclusive $e^+\,e^-$ annihilation (SIA) experiment performed by OPAL at LEP has not observed significant signal on the transverse polarization of the $\Lambda$ hyperon~\cite{Ackerstaff:1997nh}.
As an alternative to SIA, the processes $e^+e^-\rightarrow \Lambda^\uparrow+h+X$~\cite{Boer:1997mf,Wei:2014pma,Guan:2018ckx}  and semi-inclusive deep inelastic scattering (SIDIS) $\ell\,p\rightarrow\ell'+\Lambda^\uparrow+X$ have been suggested~\cite{Boer:1997mf} to study the $\Lambda$ spin asymmetry, where $D^\perp_{1T}$ contribute to the transverse polarization of $\Lambda$.
Those measurements could provide a further understanding of the origin of the sizable transverse polarization of hyperons observed in different processes~\cite{Bunce:1976yb,Heller:1978ty,Mulders:1995dh,deFlorian:1997zj,Anselmino:2000vs,Anselmino:2001js,Airapetian:2004tw,
Koike:2017fxr,Gamberg:2018fwy,
Anselmino:2019cqd}.
Recently, a nonzero transverse polarization of $\Lambda$ production in SIA and semi-inclusive $e^+e^-\rightarrow\Lambda(\bar{\Lambda})+K^{\pm}(\pi^{\pm})+X$ processes were measured by the Belle Collaboration~\cite{Guan:2018ckx}, making the extraction~\cite{DAlesio:2020wjq,Callos:2020qtu,Chen:2021hdn} of the polarizing FF of $\Lambda$ possible.
On the other hand, model calculations may also provide an approach to acquire knowledge of the Lambda polarizing FF $D_{1T}^\perp$.
A calculation of $D_{1T}^\perp$ for light flavors based on a spectator-diquark model has been performed in Ref.~\cite{Yang:2017cwi} and the result was used to make predictions on physical observables.

The main purpose of this work is to study the role of the polarizing $\Lambda$ FF $D_{1T}^\perp$  in the transverse-spin dependent $\sin(\phi_{\Lambda}-\phi_S)$ asymmetry in SIDIS.
Particularly, we apply the TMD factorization~\cite{Collins:1981uk,Collins:1984kg,Ji:2004wu,Ji:2004xq,Collins:2011zzd} to estimate the spin-dependent cross section in
$l+p \rightarrow l^\prime +\Lambda^\uparrow+X$ as well as the unpolarized cross section.
The asymmetry can be expressed as the ratio of the two cross sections.
In the last two decades, TMD factorization has been widely applied in various high energy processes~\cite{Boer:2008fr,Arnold:2008kf,Aybat:2011zv,Collins:2011zzd,Collins:2012uy,Echevarria:2012pw,
Echevarria:2012js,Pitonyak:2013dsu,Echevarria:2014xaa,Kang:2015msa,Bacchetta:2017gcc,Wang:2017zym,Wang:2018pmx,
Li:2019uhj}.
Within the TMD factorization, the differential cross section in the small transverse momentum region $P_{hT}/z_h \ll Q$~($P_{hT}$ is the transverse momentum of the final-state hadron and $Q$ is the virtuality of the photon) can be expressed as the convolution of the hard scattering factors and the well-defined TMD distributions and fragmentation functions.
In our case, the $\sin(\phi_{\Lambda}-\phi_S)$  asymmetry is contributed by the convolution of $f_1$, $D_{1T}^\perp$ and the hard scattering factors.
The TMD formalism also encodes the evolution information of TMD PDFs and FFs, governed by the so-called Collins-Soper equation~\cite{Collins:1981uk,Collins:1984kg,Collins:2011zzd,Idilbi:2004vb}.
The solution of the equation is usually expressed as an exponential form of the Sudakov-like form factor~\cite{Collins:1984kg,Collins:2011zzd,Aybat:2011zv,Collins:1999dz} which determines the scale dependence of TMDs.
Therefore, in this work, we will consider the TMD evolution effect of the polarizing FF $D_{1T}^\perp$, which is not usually included in the previous calculation for the semi-inclusive $\Lambda$ production~\cite{DAlesio:2020wjq,Callos:2020qtu,Li:2021ibn}.
In the literature, several TMD evolution formalisms have been developed~\cite{Collins:1984kg,Davies:1984sp,Ellis:1997sc,Landry:2002ix,Konychev:2005iy,
Collins:2011zzd,Aybat:2011zv,Aybat:2011ge,Kang:2011mr,Echevarria:2012js,Su:2014wpa,
Echevarria:2014xaa,Echevarria:2014rua,Bacchetta:2017gcc}.
Particularly, the nonperturbative parts of the Sudakov form factor for the TMD PDFs and FFs have been extracted from experimental data based on different parameterizations.
In this work, we will adopt two parameterizations on the nonperturbative part to estimate the asymmetry~\cite{Echevarria:2014xaa,Bacchetta:2017gcc} for comparison.

The remaining content of the paper is organized as follows.
In Sec.~\ref{Sec.formalism}, we present the formalism of the $\sin (\phi_\Lambda-\phi_S)$ asymmetry in the process $l p\to e^-\Lambda^\uparrow X$  within the TMD factorization.
In Sec.~\ref{Sec.evolution}, we investigate the evolution effect for the TMD PDFs and FFs.
Particularly, we discuss the parametrization of the nonperturbative Sudakov form factors associated with the studied TMD functions in details.
In Sec.~\ref{Sec.numerical}, we present the numerical estimate on the $\sin (\phi_\Lambda-\phi_S)$ asymmetry in the $e^-p\to e^-\Lambda^\uparrow X$ process at the kinematical region of EIC with different choices on the nonperturbative part associated with TMD evolution effect.
Finally, We summarize the paper in Sec.~\ref{Sec.conclusion}.

\section{The $\sin (\phi_\Lambda-\phi_S)$ asymmetry in the $l p\to l \Lambda^\uparrow X$ process}

\label{Sec.formalism}

The process under study is the semi-inclusive deep inelastic scattering process:
\begin{align}
l(\ell)+p(P)\rightarrow l(\ell^\prime)+\Lambda^\uparrow(P_\Lambda)+X,
\end{align}
in which the lepton beam with momentum $\ell$ scatters off an unpolarized proton target $p$ with momentum $P$.
In the final state, the scattered lepton momentum $\ell^\prime$ is measured together with a transversely polarized $\Lambda$ hyperon, with $P_\Lambda$ being the momentum of the $\Lambda$ hyperon.
We define the space-like momentum transfer $q=\ell-\ell^\prime$ and $Q^2=-q^2$.
The usual invariants in SIDIS process are introduced as
\begin{align}
x_B=\frac{Q^2}{2P\cdot q},~~~y=\frac{P\cdot q}{P\cdot \ell}=\frac{Q^2}{x_Bs},~~~z=\frac{P\cdot P_\Lambda}{P\cdot q},~~~s=(P+\ell)^2,
\end{align}
where $s$ is the total center of mass energy squared, $x_B$ is the Bjorken variable, $y$ is the inelasticity and $z$ is the momentum fraction of the final state $\Lambda$ hyperon.
The corresponding six-fold~($x_B$, $y$, $z$, $\phi_\Lambda$, $\phi_S$ and $P_{\Lambda T}^2$ ) differential cross-section~(for a transversely polarized $\Lambda$ production) in the $\gamma^*N$ collinear frame can be given as~\cite{Barone:2001sp,Barone:2010zz,Yang:2016qsf}
\begin{align}
\frac{d^{6}\sigma}{dx_B dy dz d\phi_\Lambda d\phi_S dP_{\Lambda T}^2}
=\frac{\alpha^2}{x_ByQ^2}\times \left\{(1-y+\frac{1}{2}y^2)F_{UUU}+S_\perp [\sin(\phi_\Lambda-\phi_S)(1-y+\frac{1}{2}y^2)F_{UUT}^{\sin(\phi_\Lambda-\phi_S)}]\right\}.
\label{cross-section}
\end{align}
In the $\gamma^*N$ collinear frame, the momentum direction of the virtual photon is defined as the $z$-axis, the hadron plane is determined by the $z$-axis and the momentum direction of $\Lambda$, and the lepton plane is given by $\ell$ and $\ell^\prime$.
Hence $\phi_\Lambda$ stands for the azimuthal angle between the lepton and hadron planes, while $\phi_S$ is the azimuthal angle of the transverse spin vector of $\Lambda$, $\bm{P}_{\Lambda T}$ is the component of $P_\Lambda$ transverse to $\bm q$ with $\bm{P}_{\Lambda T}=-zq_T$~\cite{Boer:2011fh}.
Here, $P_{\Lambda T}$ is the characteristic transverse momentum detected in the SIDIS process, the value of which determines the validity of TMD factorization, i.e., if $ P_{\Lambda T}^2/z^2\ll Q^2$, TMD factorization can be applied and the process is sensitive to the TMD PDFs/FFs~\cite{Collins:2011zzd}.
$F_{UUU}$ and $F_{UUT}^{\sin(\phi_\Lambda-\phi_S)}$ are the spin-averaged and transverse spin-dependent structure functions, with the first, second and third subscripts denoting the polarization of the lepton beam, proton target and the final-state hadron~($\Lambda$ hyperon), respectively~(U=unpolarized, L=longitudinally polarized, T=transversely polarized).

We can define the single transverse-spin asymmetry with a $\sin (\phi_\Lambda-\phi_S)$ modulation as following~\cite{Barone:2010zz}
\begin{align}
A_{UUT}^{\sin (\phi_\Lambda-\phi_S)}(x_B,y,z,P_{\Lambda T}^2)=\frac{\frac{1}{xyQ^2}(1-y+\frac{1}{2}y^2)F_{UUT}^{\sin(\phi_\Lambda-\phi_S)}}
{\frac{1}{xyQ^2}(1-y+\frac{1}{2}y^2)F_{UUU}}.
\label{A_UUT}
\end{align}
According to TMD factorization, the structure functions $F_{UUU}$ and  $F_{UUT}^{\sin(\phi_\Lambda-\phi_S)}$ can be expressed as the convolution of the TMD PDF and TMD FF as~\cite{Barone:2010zz,Yang:2016qsf}
\begin{align}
F_{UUU}(Q;P_{\Lambda T}^2)=&\mathcal{C}[f_1D_1],
\label{F_UU}\\
F_{UUT}^{\sin(\phi_\Lambda-\phi_S)}(Q;P_{\Lambda T}^2)=&\mathcal{C}[\frac{\bm{\hat{h}}\cdot\bm{k}_{T}}{M_\Lambda}f_1D_{1T}^\perp ],
\label{F_UUT}
\end{align}
where the unit vector $\bm{\hat{h}}$ is defined as $\bm{\hat{h}}=\frac{\bm{P}_{\Lambda T}}{P_{\Lambda T}}$~\cite{Boer:1999mm,Arnold:2008kf}, and the notation $\mathcal{C}$ denotes the convolution of the transverse momenta
\begin{align}
&\mathcal{C}[\omega fD]= x\sum_{q}e_q^2\int d^2\bm{p}_T d^2\bm{k}_T\delta^2(\bm{p}_T-\bm{k}_T+\bm{q}_T)\omega(\bm{p}_T,\bm{k}_T)f^{q}(x,\bm{p}_T^2)D^{q}(z,\bm{k}_T^2),
\label{convolution}
\end{align}
with $\omega(\bm{p}_T,\bm{k}_T)$ being an arbitrary function of $\bm{p}_T$ and $\bm{k}_T$. $f_1(x_B,\bm{p}_T^2)$ is the unpolarized TMD PDF of the proton. $D_1(z,\bm{k}_T^2)$ and $D_{1T}^\perp(z,\bm{k}_T^2)$ are the unpolarized FF and the transversely polarizing FF.
The transverse momentum $\bm{k}_T$ is related to the transverse momentum of the produced hadron with respect to the quark through $\bm{K}_\perp=-z\bm{k}_T$.

It is convenient to deal with the TMD evolution effect in the $\bm b_\perp$ space that is conjugate to the transverse momentum space through Fourier transformation, since it can turn the complicated convolution in the transverse momentum space into simple product.
Therefore, we perform a transformation for the delta function
\begin{align}
\delta^{2}(\bm{p}_T-\bm{k}_T+\bm{q}_T) = {1\over (2\pi)^2}\int d^2 \bm{b}_\perp e^{-i \bm b_\perp\cdot(
\bm{p}_T-\bm{k}_T+\bm{q}_T)},
\end{align}
and obtain the following explicit form of the spin-averaged structure function $F_{UUU}$
\begin{align}
\mathcal{C}[f_1D_1]
=&x\sum_q e_q^2 \int d^2\bm{p}_T d^2\bm{k}_T \delta^2(\bm{p}_T-\bm{k}_T+\bm{q}_{T})f_1^{q/p}(x_B,\bm{p}_T^2;Q)D_1^{\Lambda/q}(z,\bm{k}_T^2;Q)\nonumber\\
=&x\frac{1}{z^2}\sum_{q}e_q^2\int d^2\bm{p}_T d^2\bm{K}_\perp
\delta^2(\bm{p}_T+{\bm{K}_\perp}/{z}-{\bm{P}_{\Lambda T}}/{z})
f_1^{q/p}(x_B,\bm{p}_T^2;Q)D_1^{\Lambda/q}(z,\bm{K}_\perp^2;Q)\nonumber\\
=&x\frac{1}{z^2}\sum_{q}e_q^2\int \frac{d^2\bm{b}_\perp}{(2\pi)^2}
e^{i\bm{P}_{\Lambda T}\cdot\bm{b}_\perp/z}\tilde{f}_1^{q/p}(x_B,\bm{b}_\perp;Q)\tilde{D}_1^{\Lambda/q}(z,\bm{b}_\perp;Q)\nonumber\\
=&x\frac{1}{z^2}\sum_{q}e_q^2\int \frac{dbb}{2\pi}J_0(P_{\Lambda T} b/z)
\tilde{f}_1^{q/p}(x_B,b;Q)\tilde{D}_1^{\Lambda/q}(z,b;Q),
\label{eq:FUUU}
\end{align}
with $P_{\Lambda T}=|\bm{P}_{\Lambda T}|$, $b=|\bm{b}_\perp|$, $J_0$ the zeroth-order Bessel function of the first kind.
The unpolarized PDF and FF in $\bm{b}_\perp$ space can be defined as~(hereafter the tilde terms represent the ones in $\bm{b}_\perp$ space)
\begin{align}
\tilde{f}_1^{q/p}(x_B,\bm{b}_\perp;Q)&=\int{d^2\bm{p}_T e^{-i\bm{p}_T \cdot \bm{b}_\perp} f_1^{q/p}(x_B,\bm{p}_T^2;Q)},\nonumber\\
\tilde{D}_1^{\Lambda/q}(z,\bm{b}_\perp;Q)&=\int{d^2\bm{K}_\perp e^{-i\bm{K}_\perp \cdot \bm{b}_\perp/z}
D_1^{\Lambda/q}(z,\bm{K}_\perp^2;Q)}.
\label{eq:Upolarizing FF}
\end{align}
Similarly, the transverse spin-dependent structure function $F_{UUT}^{\sin(\phi_\Lambda-\phi_S)}$ can be written as
\begin{align}
&\mathcal{C}[\frac{\bm{\hat{h}}\cdot\bm{k}_T}{M_\Lambda}f_1D_{1T}^\perp]\nonumber\\
=&x\sum_q e_q^2 \int{d^2\bm{p}_T d^2\bm{k}_T \delta^2(\bm{p}_T-\bm{k}_T+\bm{q}_T)\frac{\bm{\hat{h}}\cdot\bm{k}_T}{M_\Lambda}
f_1^{q/p}(x_B,\bm{p}_T^2;Q)D_{1T}^{\perp \Lambda/q}(z,\bm{k}_T^2;Q)}\nonumber\\
=&-x\frac{1}{z^3}\sum_q e_q^2 \int d^2\bm{p}_T d^2\bm{K}_\perp
\int \frac{d^2\bm{b}_\perp}{(2\pi)^2} e^{-i(\bm{p}_T+{\bm{K}_\perp}/z-{\bm{P}_{\Lambda T}}/z)\cdot \bm{b}_\perp}
(\frac{\bm{\hat{h}}\cdot\bm{K}_\perp}{M_\Lambda})
f_1^{q/p}(x_B,\bm{p}_T^2;Q)D_{1T}^{\perp \Lambda/q}(z,\bm{K}_\perp^2;Q)\nonumber\\
=&-x\frac{1}{z^3}\sum_q e_q^2
\int \frac{d^2\bm{b}_\perp}{(2\pi)^2} e^{i{\bm{P}_{\Lambda T}}\cdot \bm{b}_\perp /z}
\bm{\hat{h}}_\alpha
\tilde{f}_1^{q/p}(x_B,\bm{b}_\perp;Q)\tilde{D}_{1T}^{\perp (\alpha) \Lambda/q}(z,\bm{b}_\perp;Q),
\label{eq:FUUT}
\end{align}
where the polarizing FF of $\Lambda$ hyperon in $\bm{b}_\perp$ space is defined as
\begin{align}
\tilde{D}_{1T}^{(\alpha)\perp\Lambda/q}(z,\bm{b}_\perp;Q)&=\int{d^2\bm{K}_\perp e^{-i\bm{K}_\perp\cdot \bm{b}_\perp/z} \frac{\bm{K}_\perp^\alpha}{M_\Lambda} D_{1T}^{\perp\Lambda/q}(z,\bm{K}_\perp^2;Q)}.
\label{eq:polarizing FF}
\end{align}
The energy dependence of the TMDs here will be discussed in details in the following section.

\section{The evolution of TMD PDFs and FFs}
\label{Sec.evolution}

In this section, we set up the formalism of the TMD evolution for the unpolarized TMD PDF $f_1$ of the proton, the unpolarized TMD FF $D_1$ as well as the polarizing FF $D_{1T}^\perp$ of the $\Lambda$ hyperon.
As mentioned in the previous section, it is more convenient to express the differential cross section of the process in the $\bm b_\perp$ space than in the transverse momentum space.
Therefore, the TMD evolution of these TMDs is usually performed in the $\bm b_\perp$ space.
Generally, the TMDs $\tilde{F}(x,b;\mu,\zeta_F)$ and $\tilde{D}(z,b;\mu,\zeta_D)$ depend on two energy scales~\cite{Collins:1981uk,Collins:1984kg,Collins:2011zzd,Aybat:2011zv,Aybat:2011ge,Echevarria:2012pw}.
One is the renormalization scale $\mu$ related to the corresponding collinear PDFs or FFs, the other one is the energy scale $\zeta_F$ (or $\zeta_D$) used as a cutoff to regularize the light-cone singularity in the operator definition of TMD PDFs and FFs.
The $\zeta$-dependence is encoded in the Collins-Soper~(CS) equation as
\begin{align}
\frac{\partial\ \mathrm{ln} \tilde{F}(x,b;\mu,\zeta_F)}{\partial\ \sqrt{\zeta_F}}=\frac{\partial\ \mathrm{ln} \tilde{D}(z,b;\mu,\zeta_D)}{\partial\ \sqrt{\zeta_D}}=\tilde{K}(b;\mu),
\end{align}
with $\tilde{K}$ being the CS evolution kernel which can be computed perturbatively for small values of $b$.
The $\mu$ dependence is driven by the renormalization group equation as
\begin{align}
&\frac{d\ \tilde{K}}{d\ \mathrm{ln}\mu}=-\gamma_K(\alpha_s(\mu)),\\
&\frac{d\ \mathrm{ln} \tilde{F}(x,b;\mu,\zeta_F)}
{d\ \mathrm{ln}\mu}=\gamma_F(\alpha_s(\mu);{\frac{\zeta^2_F}{\mu^2}}),\\
&\frac{d\ \mathrm{ln} \tilde{D}(z,b;\mu,\zeta_D)}
{d\ \mathrm{ln}\mu}=\gamma_D(\alpha_s(\mu);{\frac{\zeta^2_D}{\mu^2}}),
\end{align}
where $\gamma_K$, $\gamma_F$ and $\gamma_D$ are the anomalous dimensions of $\tilde{K}$, $\tilde{F}$ and $\tilde{D}$, respectively; $\alpha_s$ is the strong coupling at the energy scale $\mu$.
More specifically, the energy evolution of the TMDs $\tilde{F}$ and $\tilde{D}$ is encoded in the Sudakov-like form factor $S$ by the exponential form $e^{-S(Q,b)}$~\cite{Idilbi:2004vb,Collins:1981uk,Collins:1984kg,Collins:2011zzd,Ji:2004wu,Collins:2014jpa,Ji:2004xq}:
\begin{align}
\tilde{F}(x,b;Q)=\mathcal{F}(Q)\times e^{-S(Q,b)}\times \tilde{F}(x,b;\mu_i),\\
\tilde{D}(z,b;Q)=\mathcal{D}(Q)\times e^{-S(Q,b)}\times \tilde{D}(z,b;\mu_i),
\label{eq:Sudakov factor}
\end{align}
where $\mathcal{F}(Q)$ and $\mathcal{D}(Q)$ are the hard factors related to the hard scattering.
For simplicity, hereafter we will set $\mu = \sqrt{\zeta_F} = \sqrt{\zeta_D}$.

It is of fundamental importance to study the TMDs in the entire $\bm b_\perp$ space, since the $b$-dependence of the TMDs can provide very useful information regarding the transverse momentum dependence of the hadronic tree-dimensional structure through Fourier transformation.
In the small $b$ region $1/Q \ll b \ll 1/ \Lambda$, the $b$-dependence is perturbatively calculable, while it turns non-perturbative in the large $b$ region.
To access the information of $\tilde{F}(x,b;\mu)$ and $\tilde{D}(x,b;\mu)$ in the entire $b$ region,
a matching procedure must be introduced with a parameter $b_{\textrm{max}}$ serving as the boundary between the two
regions. The prescription also allows for a smooth transition from perturbative to nonperturbative regions and avoids the Landau pole singularity in $\alpha_s(\mu)$.
There are different choices on the $b_\ast$-prescription in the literature.
A frequently used one is the Collins-Soper-Sterman (CSS) prescription~\cite{Collins:1984kg}
\begin{align}
\label{eq:b*}
b_\ast=b/\sqrt{1+b^2/b_{\rm max}^2}  \ ,~b_{\rm max}<1/\Lambda_\mathrm{QCD},
\end{align}
which guarantees the feature that $b_\ast\approx b$ at small $b$ value and $b_{\ast}\approx b_{\mathrm{max}}$ at large $b$ value.
To ensure that $b_\ast$ is always in the perturbative region, the typical value of $b_{\mathrm{max}}$ is chosen around 1 GeV$^{-1}$.

With the $b_\ast$ prescription in the Sudakov form factor, one can separate the Sudakov-like form factor into the perturbative part and the nonperturbative part as
\begin{equation}
\label{eq:S}
S(Q,b)=S_{\mathrm{P}}(Q,b_\ast)+S_{\mathrm{NP}}(Q,b).
\end{equation}
The perturbative part $S_{\mathrm{P}}(Q,b_\ast)$ has been studied~\cite{Echevarria:2014xaa,Kang:2011mr,Aybat:2011ge,Echevarria:2012pw,Echevarria:2014rua} in details, and has the same result for different TMD PDF and FFs:
\begin{equation}
\label{eq:Spert}
S_{\mathrm{P}}(Q,b_\ast)=\int^{Q^2}_{\mu_b^2}\frac{d\bar{\mu}^2}{\bar{\mu}^2}\left[A(\alpha_s(\bar{\mu}))
\mathrm{ln}\frac{Q^2}{\bar{\mu}^2}+B(\alpha_s(\bar{\mu}))\right],
\end{equation}
where the $A$- and $B$-coefficients in Eq.~(\ref{eq:Spert}) can be expanded as perturbative series of $\alpha_s/\pi$:
\begin{align}
A=\sum_{n=1}^{\infty}A^{(n)}(\frac{\alpha_s}{\pi})^n,\\
B=\sum_{n=1}^{\infty}B^{(n)}(\frac{\alpha_s}{\pi})^n.
\end{align}
In this work, we will take $A^{(n)}$ up to $A^{(2)}$ and $B^{(n)}$ up to $B^{(1)}$ in the accuracy of next-to-leading-logarithmic (NLL) order~\cite{Collins:1984kg,Landry:2002ix,Qiu:2000ga,Kang:2011mr,Aybat:2011zv,Echevarria:2012pw} :
\begin{align}
A^{(1)}&=C_F,\\
A^{(2)}&=\frac{C_F}{2}\left[C_A\left(\frac{67}{18}-\frac{\pi^2}{6}\right)-\frac{10}{9}T_Rn_f\right],\\
B^{(1)}&=-\frac{3}{2}C_F.
\end{align}

In the perturbative region $1/Q \ll b \ll 1/ \Lambda$, other important elements are TMD PDFs and FFs at a fixed scale ($\tilde{F}(x,b;\mu)$ and $\tilde{D}(x,b;\mu)$), which can be expressed as the convolution of the perturbatively calculable coefficients $C$ and the corresponding collinear counterparts of TMDs ($F_{i/H}(\xi,\mu)$ and $D_{H/j}(\xi,\mu)$),
\begin{align}
\tilde{F}(x,b;\mu)=\sum_i \int_{x}^1\frac{d\xi}{\xi} C_{q\leftarrow i}(x/\xi,b;\mu)F_{i/H}(\xi,\mu),\\
\tilde{D}(z,b;\mu)=\sum_j \int_{z}^1\frac{d\xi}{\xi} C_{j\leftarrow q}(z/\xi,b;\mu)D_{H/j}(\xi,\mu),
\label{eq:small_b_F}
\end{align}
Here, $\mu$ is a dynamic scale related to $b_\ast$ by $\mu=c/b_\ast$ , with $c=2e^{-\gamma_E}$ and $\gamma_E\approx0.577$ being the Euler's constant~\cite{Collins:1981uk}, $C_{q\leftarrow i}(x/\xi,b;\mu)=\sum_{n=0}^{\infty}C_{q\leftarrow i}^{(n)}(\alpha_s/\pi)^n$ and $C_{j\leftarrow q}(z/\xi,b;\mu)=\sum_{n=0}^{\infty}C_{j\leftarrow q}^{(n)}(\alpha_s/\pi)^n$ are the perturatively calculable coefficient function.

The non-perturbative part $S_\mathrm{NP}$ can not be calculated from perturbative QCD. They may be parameterized and extracted from experimental data.
There are several different parametrizations on $S_\mathrm{NP}$ in the literature~\cite{Collins:1984kg,Davies:1984sp,Ellis:1997sc,Landry:2002ix,Konychev:2005iy,
Collins:2011zzd,Aybat:2011zv,Aybat:2011ge,Kang:2011mr,Echevarria:2012js,Su:2014wpa,
Echevarria:2014xaa,Echevarria:2014rua,Bacchetta:2017gcc},
we will discuss two of them in details.

\subsection{Approach I}

One of the approaches applied in this study is the Echevarria-Idilbi-Kang-Vitev (EIKV parametrization) non-perturbative Sudakov $S_\mathrm{NP}$ flor the unpolarized TMD PDFs and TMD FFs, which has the following form~\cite{Echevarria:2014xaa}:
\begin{equation}
S^\mathrm{pdf}_\mathrm{NP}(b,Q)=b^2(g_1^\mathrm{pdf}+\frac{g_2}{2}\ln{\frac{Q}{Q_0}}), \label{eq:eikv1}
\end{equation}
\begin{equation}
S^\mathrm{ff}_\mathrm{NP}(b,Q)=b^2(g_1^\mathrm{ff}+\frac{g_2}{2}\ln{\frac{Q}{Q_0}}).\label{eq:eikv2}
\end{equation}
Here, $g_2$ includes the information on the large $b$ behavior of the evolution kernel $\tilde{K}$.
This function is universal for different types of TMDs and is spin independent~\cite{Aybat:2011zv,Collins:2011zzd,Echevarria:2014xaa,Kang:2015msa}.
On the other hand, $g_1$ contains information on the intrinsic nonperturbative transverse motion of bound partons.
It could depend on the type of TMDs, and can be interpreted as the intrinsic transverse momentum width for the relevant TMDs at the initial scale $Q_0$~\cite{Aybat:2011zv,Qiu:2000ga,Qiu:2000hf,Anselmino:2012aa,Su:2014wpa}.
Furthermore, $g_1^\mathrm{pdf}$ and $g_1^\mathrm{ff}$ are parameterized as:
\begin{equation}
g_1^\mathrm{pdf}=\frac{\langle k_T^2\rangle_{Q_0}}{4},
\end{equation}
\begin{equation}
g_1^\mathrm{ff}=\frac{\langle p_T^2\rangle_{Q_0}}{4z^2},
\end{equation}
where $\langle k_T^2\rangle_{Q_0}$ and $\langle p_T^2\rangle_{Q_0}$ are the averaged intrinsic transverse momenta squared for TMD PDFs and FFs at the initial scale $Q_0$, respectively.
In Ref.~\cite{Echevarria:2014xaa} the authors tuned the current extracted ranges of three parameters $\langle k_T^2\rangle_{Q_0}, \langle p_T^2\rangle_{Q_0}$ and $g_2$ with $Q_0=\sqrt{2.4}\ \textrm{GeV}$ in
Refs.~\cite{Anselmino2005,Collins:2005ie,Schweitzer:2010tt} and further found that the following values of parameters can reasonably describe the SIDIS data together with the Drell-Yan lepton pair and $W/Z$ boson production data:
\begin{equation}
\langle k_T^2\rangle_{Q_0}=0.38\ \textrm{GeV}^2, ~~~~~~\langle p_T^2\rangle_{Q_0}=0.19\ \textrm{GeV}^2, ~~~~~~g_2=0.16\ \textrm{GeV}^2,~~~~~b_\mathrm{max}=1.5\ \textrm{GeV}^{-1}.
\end{equation}
Since the information of the nonperturbative Sudakov form factor for the polarizing FF of the $\Lambda$ hyperon still remains unknown, we assume it to be the same as that for the unpolarized TMDFF $S^\mathrm{ff}_\mathrm{NP}$.

It is straightforward to rewrite the scale-dependent TMDs $\tilde{F}$ and $\tilde{D}$ in $\bm b_\perp$ space
\begin{align}
\label{eq:tildeF}
\tilde{F}_{q/H}(x,b;Q)=e^{-\frac{1}{2}S_{\mathrm{P}}(Q,b_\ast)-S^{F_{q/H}}_{\mathrm{NP}}(Q,b)}F_{q/H}(x,\mu),\\
\tilde{D}_{H/q}(z,b;Q)=e^{-\frac{1}{2}S_{\mathrm{P}}(Q,b_\ast)-S^{D_{H/q}}_{\mathrm{NP}}(Q,b)}D_{H/q}(z,\mu)
\end{align}
Hereafter, we apply the leading order (LO) results for the hard coefficients $C$, $\mathcal{F}$ and $\mathcal{D}$ for $f_1$, $D_1$ and $D_{1T}^\perp$, i.e. $C_{q\leftarrow i}^{(0)}=\delta_{iq}\delta(1-x)$, $C_{j\leftarrow q}^{(0)}=\delta_{qj}\delta(1-z)$, $\mathcal{F}(Q)=1$ and $\mathcal{D}(Q)=1$.
The factor of $\frac{1}{2}$ in front of $S_{\mathrm{P}}$ comes from the fact that $S_{\mathrm{P}}$ is equally distributed to the initial-state quark and the final-state quark~\cite{Prokudin:2015ysa}.

With all the above ingredients, we can write down the evolved TMDs explicitly as
\begin{align}
\tilde{f}_1^{q/p}(x_B,b;Q) &=e^{-\frac{1}{2}S_{\mathrm{P}}(Q,b_\ast)-S^{pdf}_{\mathrm{NP}}(Q,b)}
 f_1^{q/p}(x_B,\mu),\label{eq:f_b}\\
 \tilde{D}_1^{\Lambda/q}(z,b;Q) &=e^{-\frac{1}{2}S_{\mathrm{P}}(Q,b_\ast)-S^{ff}_{\mathrm{NP}}(Q,b)}
 D_1^{\Lambda/q}(z,\mu),\label{eq:D_b}\\
\tilde{D}_{1T}^{\perp (\alpha) \Lambda/q}(z,b;Q)
&=\frac{i\bm{b}_\perp^\alpha}{2}e^{-\frac{1}{2}S_{\mathrm{P}}(Q,b_\ast)-S^{ff}_{\mathrm{NP}}(Q,b)}
 \hat{D}_{1T}^{\perp(3)}(z,z,\mu).\label{eq:D_1Tb}
\end{align}
Here, $\hat{D}_{1T}^{\perp(3)}(z,z,\mu)$ is the twist-3 FF of quark flavor $q$ to $\Lambda$ hyperon, which satisfies the following relation with the polarizing FF $D_{1T}^{\perp \Lambda/q}$ and the first transverse moment of the polarizing FF $D_{1T}^{\perp(1)}$ ~\cite{Yuan:2009dw}:
\begin{align}
\hat{D}_{1T}^{\perp(3)}(z,z,\mu)=\int{d^2\bm{K}_\perp\frac{|\bm{K}_\perp^2|}{M_\Lambda}D_{1T}^{\perp h/q}(z,\bm{K}_\perp^2,\mu)}=2M_{\Lambda}D_{1T}^{\perp(1)}(z,\mu).
\label{twist-3 polarizing FF}
\end{align}

Thus, the TMDs in the transverse momentum space can be obtained by performing the Fourier transformation
\begin{align}
f_1^{q/p}(x_B,p_T;Q)&=\int_0^\infty\frac{db b}{2\pi}J_0(p_T b)e^{-\frac{1}{2}S_{P}(Q,b_\ast)-S^\mathrm{pdf}_{\mathrm{NP}}(Q,b)} f_1^{q/p}(x_B,\mu),\label{eq:f-D1pion}\\
D_1^{\Lambda/q}(z,K_\perp;Q)&=\int_0^\infty\frac{db b}{2\pi}J_0(K_\perp b/z)e^{-\frac{1}{2}S_{P}(Q,b_\ast)-S^\mathrm{ff}_{\mathrm{NP}}(Q,b)} D_1^{\Lambda/q}(z,\mu),\label{eq:f-D1Lambda}\\
\frac{K_\perp^\alpha}{M_\Lambda}\tilde{D}_{1T}^{\perp \Lambda/q(\alpha)}(z,K_\perp;Q)&=\int_0^\infty\frac{db b^2}{4\pi}J_1(K_\perp b/z)e^{-\frac{1}{2}S_{P}(Q,b_\ast)-S^\mathrm{ff}_{\mathrm{NP}}(Q,b)} D_{1T}^{\perp(3)}(z,z,\mu).
\end{align}
where $p_T = |\bm p_T|$, $K_\perp = |\bm K_\perp|$.

\subsection{Approach II}

Besides the traditional parametrization~\cite{Nadolsky:1999kb,Landry:2002ix,Konychev:2005iy} and the EIKV parametrization, some other forms have been also proposed~\cite{Aidala:2014hva,Collins:2014jpa,Su:2014wpa,Bacchetta:2017gcc} recently.
Particularly, a new evolution formalism was proposed by Bacchetta, Delcarro, Pisano, Radici and Signori (BDPRS parametrization) in Ref.~\cite{Bacchetta:2017gcc} for $\tilde{f}_1^a$ and $\tilde{D}^{a\rightarrow h}_1$:
\begin{align}
\tilde{f}^{a}_1(x,b^2;Q^2)&=f^{a}_1(x;\mu^2)e^{-S(\mu^2,Q^2)}
e^{\frac{1}{2}g_K(b)\textrm{ln}(Q^2/Q_0^2)}\tilde{f}^{a}_{1\mathrm{NP}}(x,b^2),\label{eq:SNP-jhep-f}\\
\tilde{D}^{a\rightarrow h}_1(z,b^2;Q^2)&=D^{a\rightarrow h}_1(z;\mu^2)e^{-S(\mu^2,Q^2)}e^{\frac{1}{2}g_K(b)\textrm{ln}(Q^2/Q_0^2)}\tilde{D}^{a\rightarrow h}_{1\mathrm{NP}}(z,b^2),\label{eq:SNP-jhep1}
\end{align}
where $g_K=-g_2b^2/2$, following the choice in Refs.~\cite{Landry:2002ix,Nadolsky:1999kb,Konychev:2005iy}. $\tilde{f}^{a}_{1\mathrm{NP}}(x,b^2)$ and $\tilde{D}^{a\rightarrow h}_{1\mathrm{NP}}(z,b^2)$ are the intrinsic nonperturbative part of the PDFs and FFs respectively, which are parameterized as
\begin{equation}
\label{eq:SNP-jhep2-f}
\tilde{f}^a_{1\mathrm{NP}}(x,b^2)=\frac{1}{2\pi}e^{-g_{1}\frac{b^2}{4}}(1-\frac{\lambda g^2_{1}}{1+\lambda g_{1}}\frac{b^2}{4}),
\end{equation}
\begin{equation}
\label{eq:SNP-jhep2}
\tilde{D}^{a\rightarrow h}_{1\mathrm{NP}}(z,b^2)=\frac{g_3e^{-g_3\frac{b^2}{4z^2}}+(\frac{\lambda_F}{z^2})g_4^2(1-g_4\frac{b^2}{4z^2})e^{-g_4\frac{b^2}{4z^2}}}
{2\pi z^2 (g_3+(\frac{\lambda_F}{z^2})g_4^2)},
\end{equation}
with
\begin{equation}
g_{1}(x)=N_{1}\frac{(1-x)^\alpha x^\sigma}{(1-\hat{x})^\alpha \hat{x}^\sigma},
\end{equation}
\begin{equation}
g_{3,4}(z)=N_{3,4}\frac{(z^\beta+\delta)(1-z)^\gamma}{(\hat{z}^\beta+\delta)(1-\hat{z})^\gamma}.
\end{equation}
Here, $\hat{x}=0.1$ and $\hat{z}=0.5$ are fixed, and $\alpha, \sigma, \beta, \gamma, \delta$, $N_1\equiv g_1(\hat{x})$, $N_{3,4}\equiv g_{3,4}(\hat{z})$ are free parameters fitted to the available data from SIDIS, Drell-Yan, and $W/Z$ boson production processes.
Besides the $b_\ast(b)$ prescription in the original CSS approach~\cite{Collins:1984kg}, there are also several different choices on the form of $b_\ast(b)$~\cite{Collins:2016hqq,Bacchetta:2017gcc}.
In Ref.~\cite{Bacchetta:2017gcc}, a new $b_\ast$ prescription different from Eq.~(\ref{eq:b*}) was proposed as
\begin{equation}
b_\ast=b_\mathrm{max}\left(\frac{1-e^{{-b^4}/{b_\mathrm{max}^4}}}{1-e^{{-b^4}/{b_\mathrm{min}^4}}}\right)^{1/4}
\label{b*2}
\end{equation}
Again, $b_\mathrm{max}$ is the boundary of the nonperturbative and perturbative $\bm b_\perp$ space region with fixed value of $b_\mathrm{max}=2e^{-\gamma _E}\ \textrm{GeV}^{-1}\approx 1.123\ \textrm{GeV}^{-1}$.
Furthermore, the authors in Ref.~\cite{Bacchetta:2017gcc} also chose to saturate $b_\ast$ at the minimum value $b_\mathrm{min}\propto 2e^{-\gamma_E}/Q$.

In this work, we will adopt the both the EIKV evolution formalism and the BDPRS evolution formalism to estimate the $\sin(\phi_\Lambda-\phi_S)$ asymmetry in SIDIS. The goal is to investigate the impact of the different evolution formalisms on the asymmetry.

\section{Numerical calculation}

\label{Sec.numerical}

Using the framework set up above, we perform the numerical calculation on the $\sin (\phi_\Lambda-\phi_S)$ azimuthal asymmetry in the process $e^-p\to e^-\Lambda^\uparrow X$ at the kinematical region of EIC.
To do this we need to know the collinear functions appearing in Eqs.~(\ref{eq:f_b}), (\ref{eq:D_b}) and (\ref{eq:D_1Tb}).
For the unpolarized PDF $f_1(x, \mu)$ of the proton target, we apply the NLO set of the CT10 parametrization (central PDF set)~\cite{Lai:2010vv}.
For the $D_{1T}^{\perp(3)}(z,\mu)$  and $D_1(z,\mu)$ of the $\Lambda$ hypeon,
we adopt two different sets for comparison.

The first set (Set I) is the polarizing FF of lambda $D_{1T}^{\perp}$ for light flavors from the spectator diquark model calculation~\cite{Yang:2017cwi}, in which the contributions from both the scalar diquark and the axial-vector diquark spectators are included.
Assuming the SU(6) spin-flavor symmetry, the fragmentation functions of the $\Lambda$ hyperon for light flavors satisfy the relations between different quark flavors and diquark types as
\begin{align}
D_{1T}^{\perp u}=D_{1T}^{\perp d}=\frac{1}{4}D_{1T}^{\perp (s)}+\frac{3}{4}D_{1T}^{\perp (v)},~~~~~~~~~D_{1T}^{\perp s}=D_{1T}^{\perp (s)},
\label{polarizing FF-uds}
\end{align}
where $u$, $d$ and $s$ denote the up, down and strange quarks, respectively.
$D_{1T}^{\perp (v)}$ and $D_{1T}^{\perp (s)}$ represent the contribution from the axial-vector diquark and scalar diquark, and their expressions as well as the values of the parameters can be found in Refs.~\cite{Yang:2017cwi}.
One should notice that in this model only the valence quarks contribute to the $\Lambda$ fragmentation function, while the sea quark contribution is zero.
For consistency, in this set, we apply the FF $D_1^{\Lambda/q}(z)$ from the same model in Ref.~\cite{Yang:2017cwi}.
Furthermore, we apply the {\sc{QCDNUM}} package~\cite{Botje:2010ay} to perform the Dokshitzer-Gribov-Lipatov-Altarelli-Parisi (DGLAP) evolution of unpolarized FF $D_1^{\Lambda/q}(z,\mu_b)$ and
the polarizing FF $\hat{D}_{1T}^{\perp(3)}(z,z,\mu_b)$ from the model scale $Q_0$ to another energy, since the evolution kernel for the diagonal piece of $\hat{D}_{1T}^{\perp(3)}(z,z,\mu_b)$ is the same as that for the unpolarized FF~\cite{Kang:2010xv}.

\begin{figure}
  \centering
  \includegraphics[width=0.329\columnwidth]{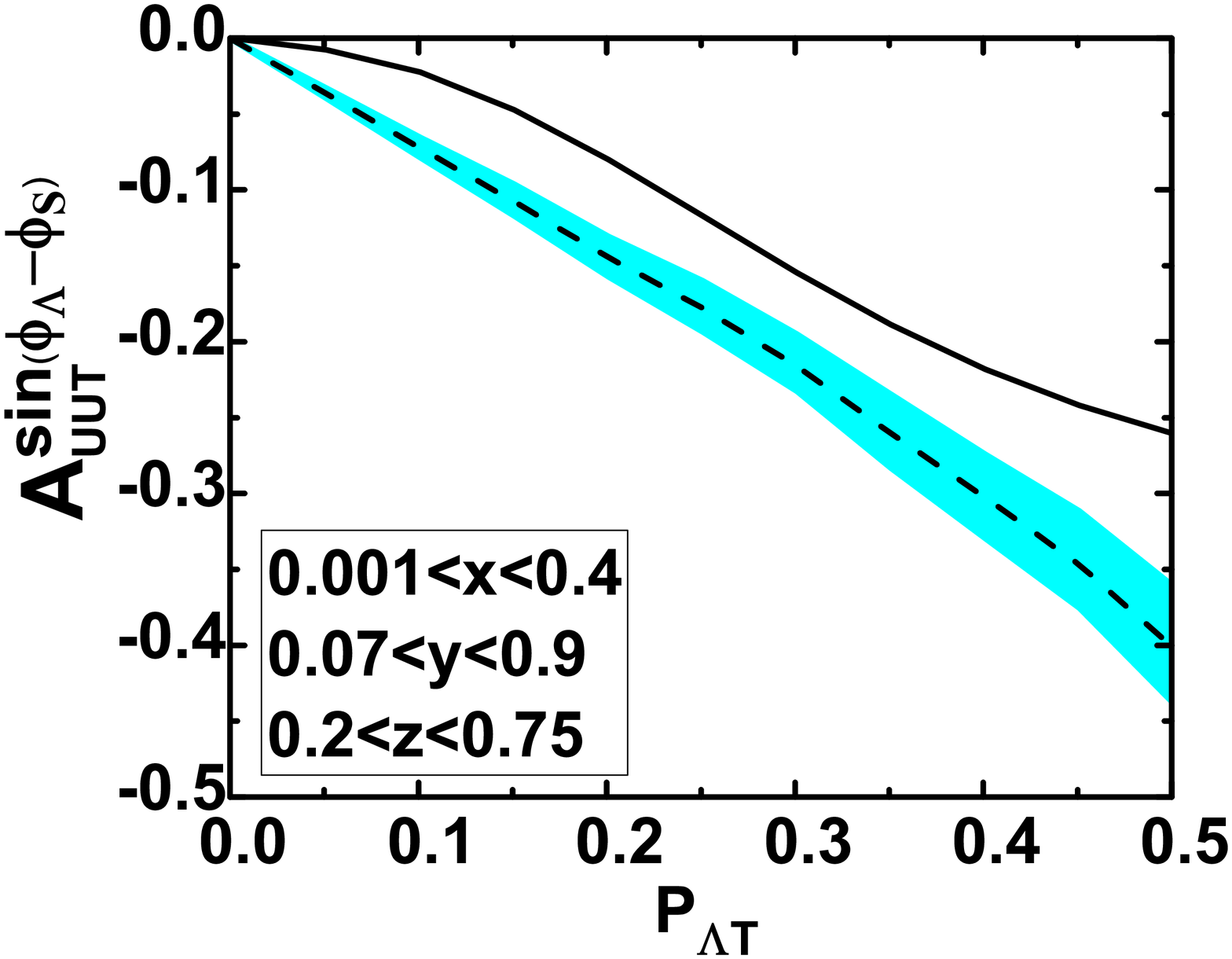}
  \includegraphics[width=0.329\columnwidth]{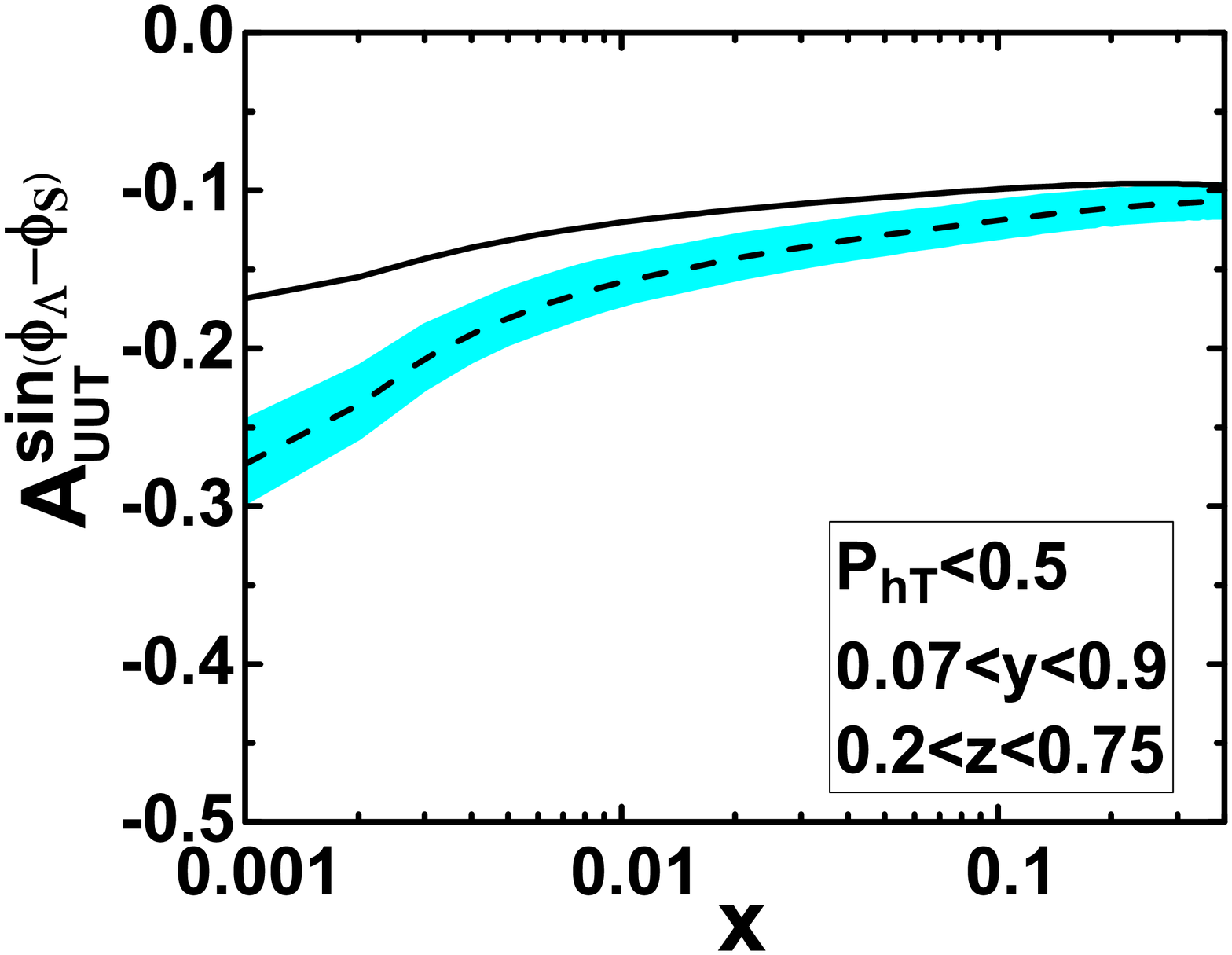}
  \includegraphics[width=0.329\columnwidth]{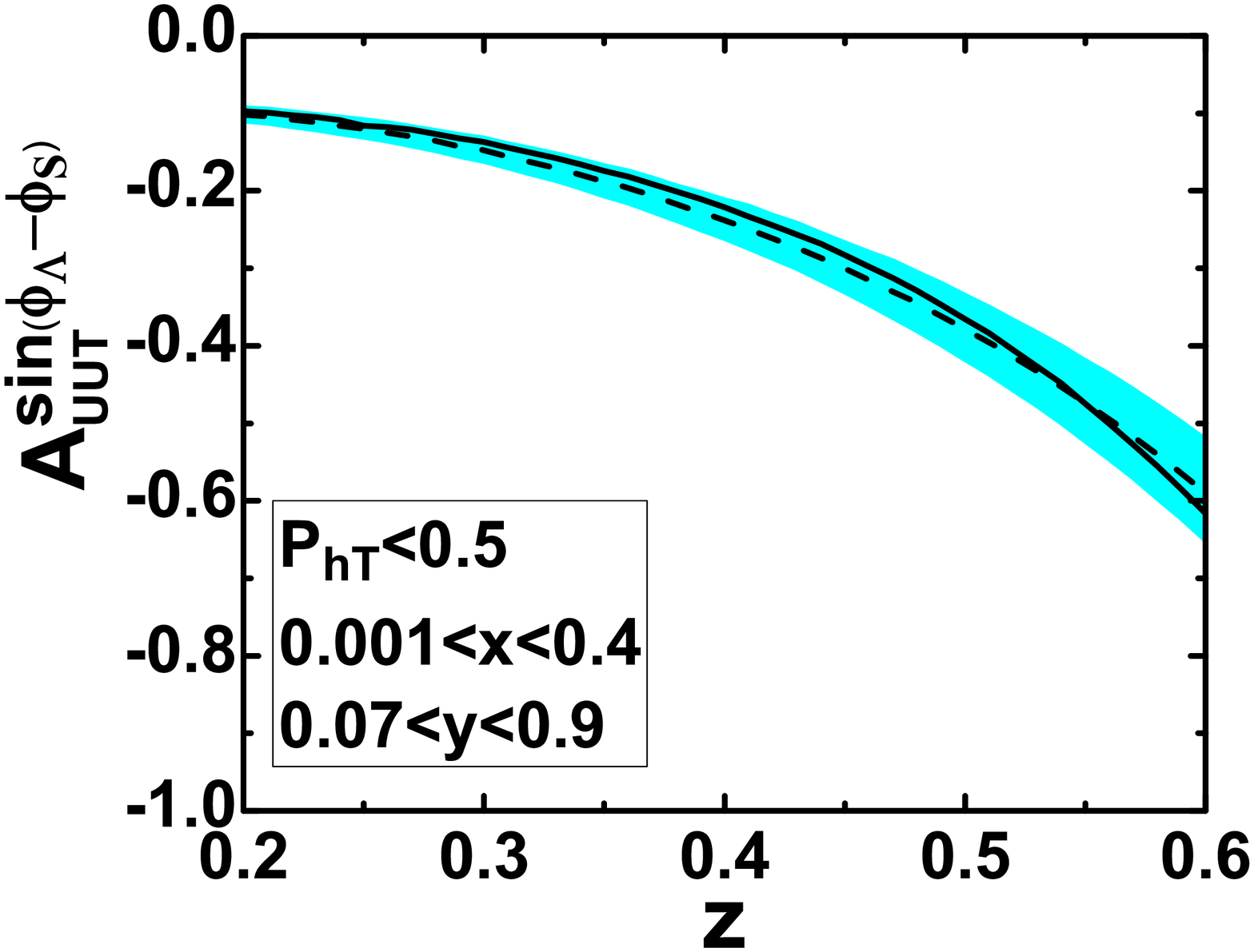}
  \caption{The $\sin (\phi_\Lambda-\phi_S)$ azimuthal asymmetry in SIDIS process $l p\rightarrow\Lambda^\uparrow+X$. The solid lines correspond to the results from the BDPRS parametrization~\cite{Bacchetta:2017gcc} [Eqs.~(\ref{eq:SNP-jhep-f}) and (\ref{eq:SNP-jhep1})] on the nonperturbative form factor, while the dashed lines correspond to the results calculated from the EIKV parametrization~\cite{Echevarria:2014xaa} [Eqs.~(\ref{eq:eikv1}) and Eq.~(\ref{eq:eikv2})]. The shaded areas show the uncertainty bands determined by the uncertainties of the parameters. In this calculation the spectator model result for the $\Lambda$ polarizing FF is adopted.}
  \label{fig1:asy-SNP}
\end{figure}

The second set (set II) of the FFs is the parametrization of the $\Lambda$ polarizing FF in Ref.~\cite{DAlesio:2020wjq}, extracted from the Belle experimental data on the $\Lambda$ /$\bar{\Lambda}$ polarization in $e^+e^-$ annihilation process, in which the inclusive (plus a jet) $\Lambda$ and associated production of a light charged hadron is measured.
In Ref.~\cite{DAlesio:2020wjq}, the first transverse-moment of $D_{1T}^{\perp}$ is given by
\begin{align}
D_{1T}^{\perp(1)}(z)=\sqrt{\frac{e}{2}}\frac{1}{zM_\Lambda}\frac{1}{M_{pol}}\frac{\langle K_\perp^2\rangle_{pol}^2}{\langle K_\perp^2\rangle}\Delta D_{\Lambda^\uparrow/q}(z),
\label{para-polarizing FF(1)}
\end{align}
with
\begin{align}
\langle K_\perp^2\rangle_{pol}=\frac{M_{pol}^2}{M_{pol}^2+\langle K_\perp^2\rangle}\langle K_\perp^2\rangle,
\label{rela}
\end{align}
where the unpolarized Gaussian width $\langle K_\perp^2\rangle=0.2\ \textrm{GeV}^2$~\cite{Anselmino2005} and the $z$-dependent part of the polarizing FF $\Delta D_{\Lambda^\uparrow/q}(z)$ was parameterized as
\begin{align}
\Delta D_{\Lambda^\uparrow/q}(z)=N_qz^{a_q}(1-z)^{b_q}\frac{(a_q+b_q)^{a_q+b_q}}{{a_q}^{a_q}{b_q}^{b_q}}
D_{\Lambda/q}(z).
\label{z-part}
\end{align}
Here, the AKK08~\cite{Albino:2008fy} set for unpolarized $\Lambda$ FF $D_{\Lambda/q}(z)$ is adopted.
Since the $\Lambda$ FF set are given for $\Lambda+\bar{\Lambda}$,
the two contributions are separated as~\cite{DAlesio:2020wjq}
\begin{align}
D_{\bar{\Lambda}/q}(z_p)=D_{\Lambda/\bar{q}}(z_p)=(1-z_p)D_{\Lambda/q}(z_p),
\label{AKK08}
\end{align}
where the scaling variable $z_p$ is related to $z$ by $z_p\simeq z[1-M_\Lambda^2/(z^2Q^2)]$.
The corresponding collinear twist-3 fragmentation function of quark flavor $q$ to $\Lambda$ hyperon $\hat{D}_{1T}^{\perp(3)}(z,z,\mu_b)$ can also be obtained by using Eq.~(\ref{twist-3 polarizing FF}).
The best fit of the parameters in Eq.~(\ref{z-part}) are obtained as
\begin{align}
N_u&=0.47^{+0.32}_{-0.20},\quad N_d=-0.32\pm0,13,\quad N_s=-0.57^{+0.29}_{-0.43},\quad a_u=0,\quad a_d=0,\nonumber\\
a_s&=2.30^{1.08}_{-0.91},\quad b_u=3.50^{+2.33}_{-1.82},\quad b_d=0,\quad b_s=0,\quad \langle K_\perp^2\rangle_{pol}=0.1\pm0.02\ \textrm{GeV}^2.
\label{canshuhua}
\end{align}

\begin{figure}
  \centering
  \includegraphics[width=0.329\columnwidth]{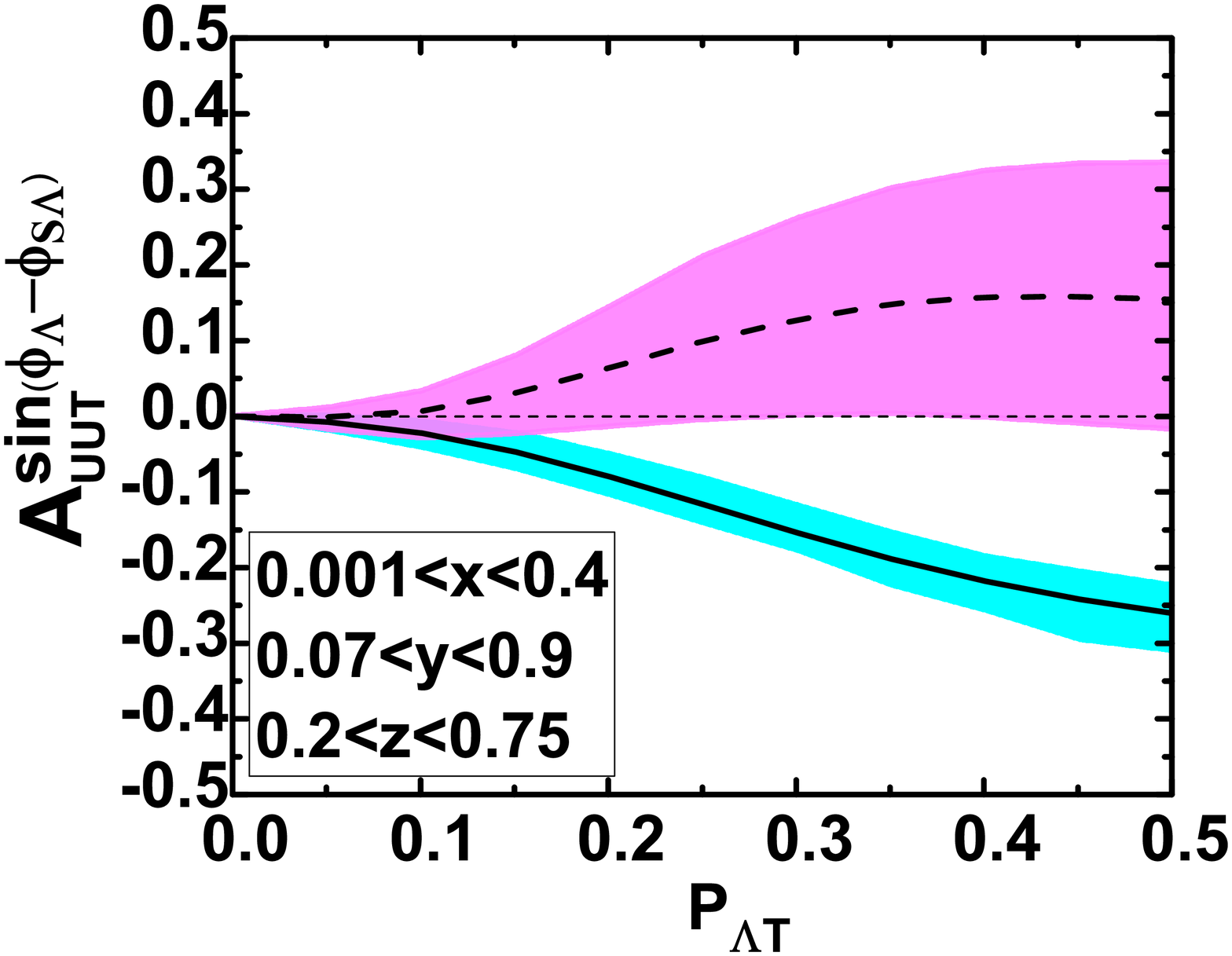}
  \includegraphics[width=0.329\columnwidth]{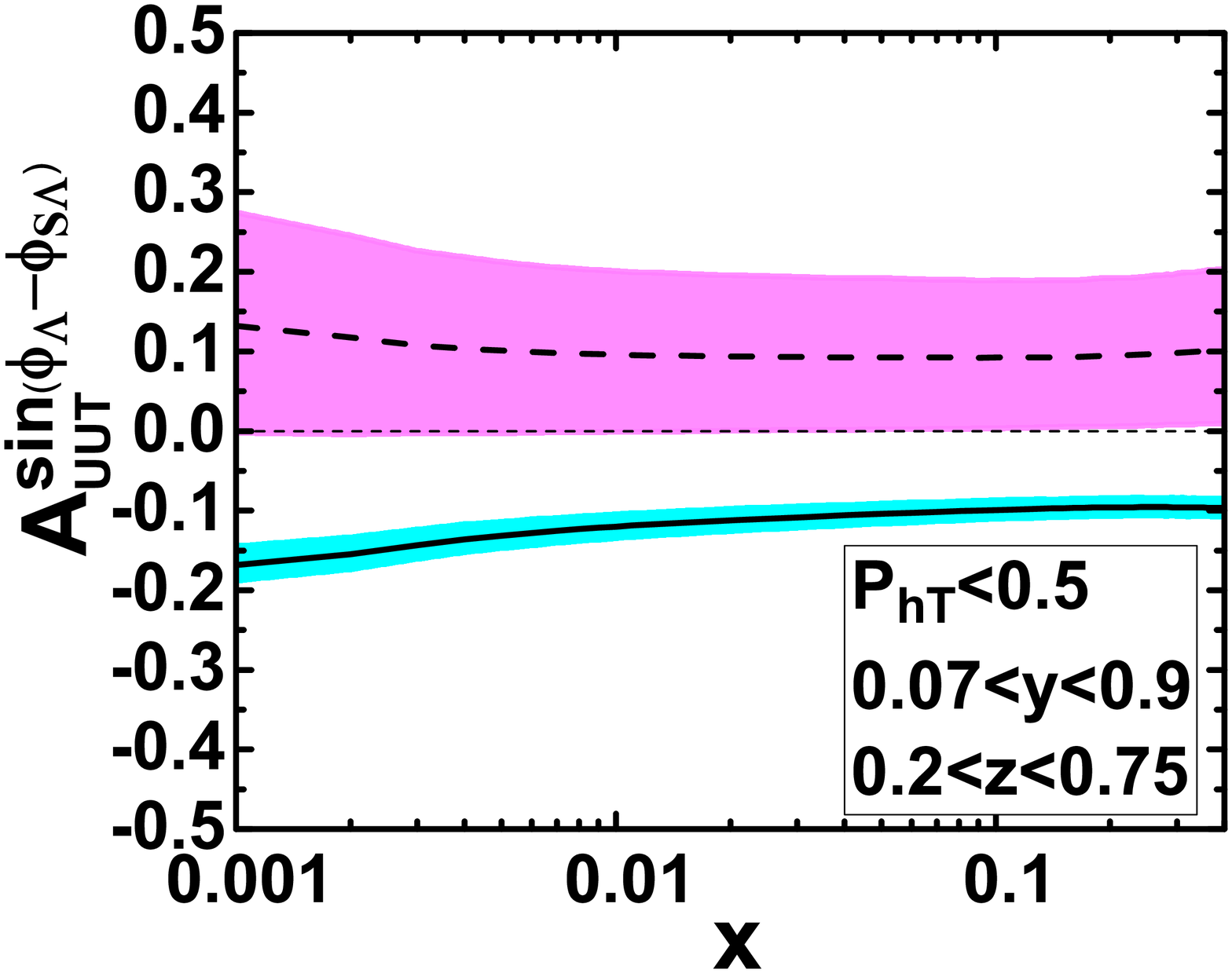}
  \includegraphics[width=0.329\columnwidth]{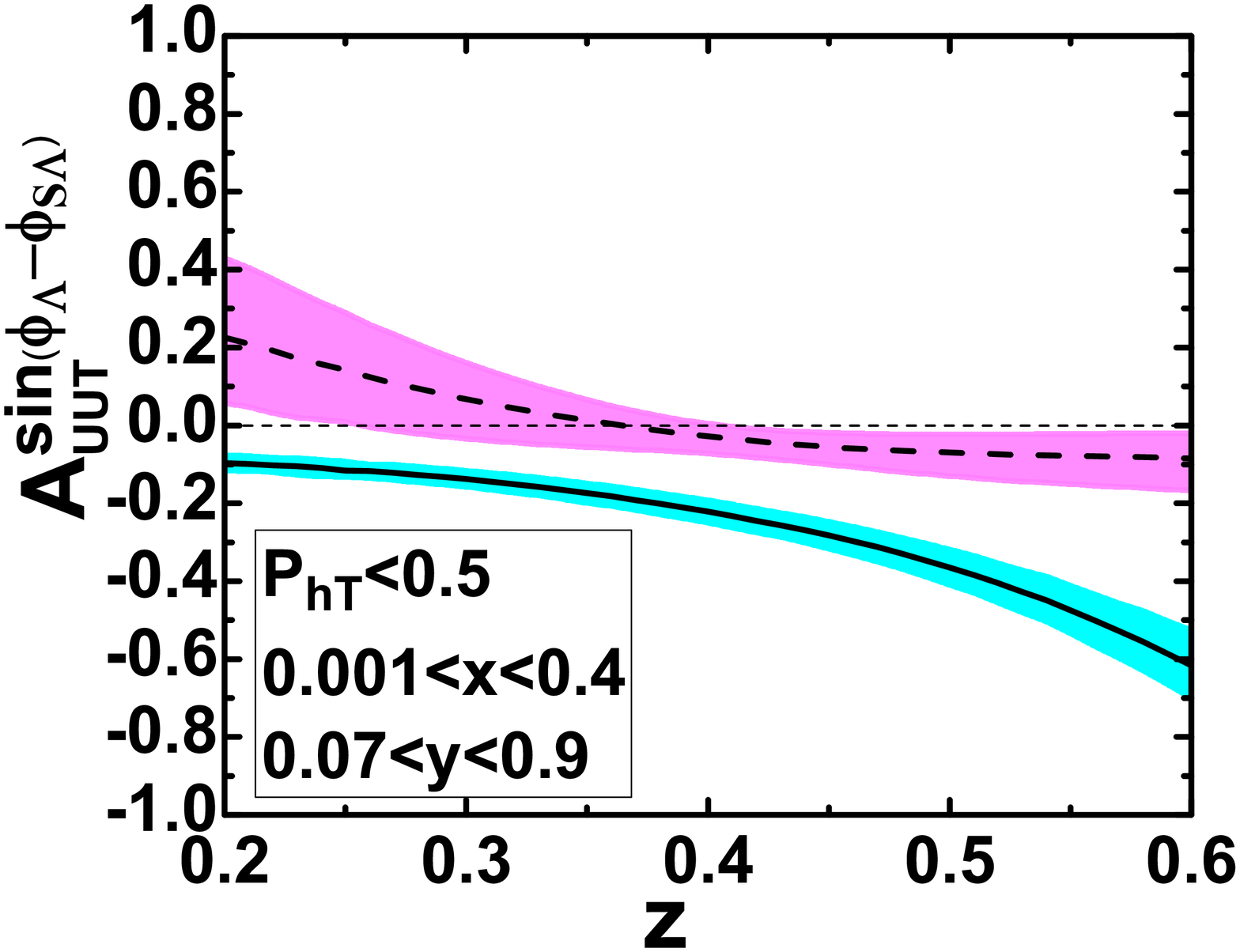}
  \caption{Comparison between the asymmetry from the set I result (spectator model result~\cite{Yang:2017cwi}) and that from the set II result (parametrization in~\cite{DAlesio:2020wjq}) for the $\Lambda$ polarizing FF, depicted by the solid lines and dashed lines, respectively.
  The shaded areas show the uncertainty bands determined by the uncertainties of the parameters. In this calculation the BDPRS parametrization~\cite{Bacchetta:2017gcc} for the TMD evolution is adopted.}
  \label{fig2:asy-polarizing FF}
\end{figure}

We apply the kinematical ranges of EIC as follows~\cite{Accardi:2012qut}
\begin{align}
&0.001<x<0.4,\quad 0.07<y<0.9,\quad 0.2<z<0.75,\nonumber\\
&1\ \textrm{GeV}^2<Q^2,\quad W>5\ \textrm{GeV},\quad \sqrt{s}=45\ \textrm{GeV},\quad P_{\Lambda T}<0.5\ \textrm{GeV},
\label{EIC}
\end{align}
with $W^2=(P+q)^2\approx\frac{1-x}{x}Q^2$ being the invariant mass of the virtual photon-nucleon system.
Using the above kinematical configurations and applying Eqs.~(\ref{A_UUT}),~(\ref{eq:FUUU}) and~(\ref{eq:FUUT}), we numerically estimate the $\sin (\phi_\Lambda-\phi_S)$ asymmetry in the electroproduction of transversely polarized $\Lambda$ at EIC.
The corresponding numerical results are plotted in Fig.~\ref{fig1:asy-SNP} and Fig.~\ref{fig2:asy-polarizing FF}, in which the left, middle, and right panels show the $\sin(\phi_\Lambda-\phi_S)$ azimuthal asymmetry as functions of $P_{\Lambda T}$, $x$ and $z$, respectively.

Fig.~\ref{fig1:asy-SNP} plots the asymmetries calculated from the spectator diquark model result (set I) of the $\Lambda$ polarizing FF~\cite{Yang:2017cwi}.
Here, two different approaches for the nonperturbative Sudakov form factor in the TMD evolution formalism are adopted for comparison.
The dashed lines correspond to the asymmetry from the EIKV parametrization~\cite{Echevarria:2014xaa} (Approach I) on $S_{\textrm{NP}}$ combined with the $b_\ast$ prescription in Eq.~(\ref{eq:b*}).
The shaded areas show the uncertainty bands due to the uncertainties of the parameters.
The solid lines show the asymmetry calculated from the BDPRS parametrization (Approach II)~\cite{Bacchetta:2017gcc} on the nonperturbative Sudakov form factor.
In this calculation, the $b_\ast$ prescription in Eq.~(\ref{b*2}) is used, which is different from the CSS prescription.
As depicted in Fig.~\ref{fig1:asy-SNP}, in all cases the $\sin(\phi_\Lambda-\phi_S)$ azimuthal asymmetries are negative and sizable.
The minus sign of the asymmetry comes from the negative results of $\Lambda$ polarizing FFs in our model calculation. In addition, the magnitude of asymmetry decreases with increasing $x$, while it increases with increasing $P_{\Lambda T}$ or $z$.
Moreover, we find that different approaches dealing with the non-perturbative part of evolution lead to the same signs and the tendencies of the asymmetries, although the size of $x$-dependent and $P_{\Lambda T}$-dependent asymmetries are somewhat different.
For the $z$-dependent asymmetry, it is found that the two approaches lead to very similar results.

To investigate the dependence of the asymmetry on different choices of the $\Lambda$ polarizing FFs, we also adopt the parametrization in (\ref{para-polarizing FF(1)})~\cite{DAlesio:2020wjq} to calculate the $\sin(\phi_\Lambda-\phi_S)$ asymmetry in SIDIS.
The results are plotted by the dashed lines in Fig.~\ref{fig2:asy-polarizing FF}).
The solid lines denote the results (solids line in Fig.~\ref{fig1:asy-SNP}) from the spectator model calculation for the $\Lambda$ polarizing FF for comparison.
In this calculation, we utilize the BDPRS parametrization (approach II) on the non-perturbative part of the TMD evolution formalism.
Again, the shaded areas in Fig.~\ref{fig2:asy-polarizing FF} show the uncertainty bands determined by the uncertainties of the parameters in the extraction of the polarizing FF.
We find that the magnitude of the $P_{\Lambda T}$- and $x$-dependent asymmetries calculated from the parametrization for $\Lambda$ polarizing FF in Ref.~\cite{DAlesio:2020wjq} is similar to the results using the $\Lambda$ polarizing FF from the spectator model, however, the sign of the asymmetry is opposite to that from the spectator model result.
Furthermore, the results calculated from the the parametrization shows a node in the $z$-dependent asymmetry.
This is because the extracted $\Lambda$ polarizing FF for the up quark in Ref.~\cite{DAlesio:2020wjq} is positive, while that for the down quark is negative.
Thus, future experimental data on the $\sin(\phi_\Lambda-\phi_S)$ asymmetry of $\Lambda$ production in SIDIS with high precision at EIC can discriminate different results for the $\Lambda$ polarizing FF.
We also note there is a relative large uncertainty band since in this calculation the errors of the parameters of the $\Lambda$ polarizing FF has also been included.

\section{Conclusion}
\label{Sec.conclusion}

In this work, we have applied the TMD factorization approach to study the $\sin (\phi_\Lambda-\phi_S)$ azimuthal asymmetry in $e^-p\to e^-\Lambda^\uparrow X$ process at the kinematical region of EIC.
The asymmetry arises from the convolution of the polarizing FF $D_{1T}^\perp$ for $\Lambda$ hyperon and the unpolarized PDF $f_1$ for the proton.
We have taken into account the TMD evolution effects of the unpolarized FF and the transversely polarizing FF $D_{1T}^\perp$ of $\Lambda$ hyperon.
In practical calculation we have taken into account two approaches for the TMD evolution for comparison. One is the EIKV approach, the other is the BDPRS approach.
Their main difference is the treatment on the nonperturbative part of evolution, while the perturbative part in the two approach are the same and have been kept at NLL accuracy in this work.
As the nonperturbative Sudakov form factor associated with the $\Lambda$ polarizing FF is still unknown, we assume that it has the same form as that of the unpolarized fragmentation function.
The hard coefficients associated with the corresponding collinear functions in the TMD evolution formalism are kept at the leading-order accuracy.
For the $\Lambda$ fragmentation at fixed scale, the model result from diquark spectator model and the extraction from Belle $e^+e^-$ data were have been adopted to estimate the asymmetry.

The numerical calculations show that different choices of nonperturbative Sudakov form factors in the TMD evolution formalism lead to similar results for $\sin (\phi_\Lambda-\phi_S)$ asymmetry at the energy scale of EIC, particularly in the $z$-dependent asymmetry.
The asymmetry utilizing the spectator model for the $\Lambda$ polarizing FF is negative in the entire kinematical regions, since the polarizing FFs of $\Lambda$ for $u$ and $d$ quarks are both negative due to the assumption of $SU(6)$ spin-flavor symmetry.
As a comparison, the $x-$dependent and $P_{\Lambda T}-$ dependent asymmetries calculated from the parametrization for the polarizing FF show positive values, and there is a node in the $z$-dependent asymmetry as the function of $z$.
Our study demonstrates that different choice on the Lambda polarizing FF can lead to very different asymmetry in SIDIS.
Future measurements on the $\sin (\phi_\Lambda-\phi_S)$ asymmetry with high precision at EIC can provide important cross check on the available $\Lambda$ polarizing FFs as well as constrain them more stringently.

\section*{Acknowledgements}
This work is partially supported by the NSFC (China) grants 11575043,11905187,11847217. X. Wang is supported by the China Postdoctoral Science Foundation under Grant No.~2018M640680 and the Academic Improvement Project of Zhengzhou University.

\end{document}